# Cross-tokamak Disruption Prediction based on Physics-Guided Feature Extraction and domain adaptation

Chengshuo Shen[1], Wei Zheng[1]*, Bihao Guo[2], Yonghua Ding[1], Dalong Chen[2], Xinkun Ai[1], Fengming Xue[1], Yu Zhong[1], Nengchao Wang[1], Biao Shen[2], Binjia Xiao[2], Zhongyong Chen[1], Yuan Pan[1] and J-TEXT team[†]

[1] State Key Laboratory of Advanced Electromagnetic Engineering and Technology, International Joint Research Laboratory of Magnetic Confinement Fusion and Plasma Physics, School of Electrical and Electronic Engineering, Huazhong University of Science and Technology, Wuhan, 430074, China

[2] Institute of Plasma Physics, Hefei Institutes of Physical Science, Chinese Academy of Sciences, Hefei 230031, China

**Email**:
shenchengshuo@hust.edu.cn, zhengwei@hust.edu.cn

**Abstract**

The high acquisition cost and the significant demand for disruptive discharges for data-driven disruption prediction models in future tokamaks pose an inherent contradiction in disruption prediction research. In this paper, we demonstrated a novel approach to predict disruption in a future tokamak using only a few discharges. The approach aims to predict disruption by finding a feature space that is universal to all tokamak. The first step is to use the existing understanding of physics to extract physics-guided features from the diagnostic signals of each tokamak, called physics-guided feature extraction (PGFE). The second step is to align a few data from the future tokamak (target domain) and a large amount of data from existing tokamak (source domain) based on a domain adaptation algorithm called CORrelation ALignment (CORAL). It is the first attempt at applying domain adaptation in the task of disruption prediction. PGFE has been successfully applied in J-TEXT to predict disruption with excellent performance. PGFE can also reduce the data volume requirements due to extracting the less device-specific features, thereby establishing a solid foundation for cross-tokamak disruption prediction. We have further improved CORAL (supervised CORAL, S-CORAL) to enhance its appropriateness in feature alignment for the disruption prediction task. To simulate the existing and future tokamak case, we selected J-TEXT as the existing tokamak and EAST as the future tokamak, which has a large gap in the ranges of plasma parameters. The utilization of the S-CORAL improves the disruption prediction performance on future tokamak. Through interpretable analysis, we discovered that the learned knowledge of the disruption prediction model through this approach exhibits more similarities to the model trained on large data volumes of future tokamak. This approach provides a light, interpretable and few data-required way by aligning features to predict disruption using small data volume from the future tokamak.

## 1. Introduction

In future tokamaks such as ITER[1], DEMO[2] and SPARC[3], disruption is considered a catastrophic event that requires reliable avoidance or mitigation[4,5,6]. Data-driven disruption prediction, benefiting from decades of data accumulation during the operation of tokamaks, is a highly feasible approach for disruption prediction. Numerous data-driven disruption predictors have been developed on JET[7–10], ASDEX-U[11], DIII-D[12,13], C-Mod[12,14], JT-60U[15], HL-2A[16,17], EAST[18–20], and J-TEXT[21–23] with high accuracy on their own tokamaks. However, the high performance operation of future tokamaks imposes a significant cost for unmitigated disruption, making it impractical to achieve large data for training such models. The large gap in device size and operation regime between future and existing tokamaks also renders using the predictors trained on existing tokamaks directly on future tokamaks less reliable. To date, there have also been many efforts and achievements in attempting to address this issue. Adaptive learning by building a predictor from scratch has also been considered to address the challenge of disruption prediction in newly deployed tokamaks, and it has yielded promising results[24–26]. Deep learning-based disruption predictors have achieved favourable results in cross-tokamak disruption prediction by mixing data from two[27] or three[28] different tokamaks. When making predictions across parameter regimes, the 'Scenario adaptive' approach has successfully utilized high-parameter data from existing tokamaks mixed with low-parameter data from the target tokamak to predict high-parameter data of the target tokamak[29].

Transfer learning[30] is a strong candidate for training cross-tokamak disruption predictors using limited target tokamak data. The current approach primarily involves training by mixing data to predict disruption in a new tokamak. However, data mixing is a fundamental method in transfer learning. There exist advanced methods within transfer learning that facilitate improved predictions of cross-tokamak disruption. Domain adaptation[31] is applicable for addressing the problem where the source and target domains have the same features and categories, but different feature distributions. Domain adaptation has been widely applied in the fields of Computer Vision (CV)[32] and Natural Language Processing (NLP)[33]. However, it is rarely mentioned in the field of magnetic confinement fusion, especially in disruption prediction. Recently, disruption prediction on EAST with different wall conditions using the domain adaptation algorithm called maximum mean discrepancy (MMD) has significantly improved the model's performance under different wall conditions[34]. Cross-tokamak disruption prediction is also a typical application scenario of domain adaptation under the assumption that the mechanism of disruption is the same in all tokamaks. Domain adaptation algorithm can be helpful in exploring a new cross-tokamak disruption prediction approach for future tokamaks.

Our team has developed a deep model for cross-tokamak disruption prediction with the application of freeze and fine-tune technique[35]. However, there is still room for improvement in the performance of cross-tokamak disruption prediction. Deep learning-based predictors are supposed to require more and diverse data for the pre-trained model to ensure generalization. To train a generalized pre-trained model for disruption prediction, relying solely on data from a single tokamak such as J-TEXT is not sufficient. Deep learning-based predictors are also naturally difficult to understand due to the complexity inherent in deep learning. Compared to deep learning-based disruption predictors, decision tree-based disruption predictors are more interpretable, required fewer data and consume fewer computational resources. However, decision tree-based models rely more on expert experience and knowledge for selecting and processing input features. The detailed analysis of input features is also a crucial factor for achieving impressive performance on the Classification And Regression Trees (CART) based adaptive predictors[26]. Extracting features through expert knowledge can partially align the data distribution among tokamaks. An interpretable disruption predictor based on physics-guided feature extraction (IDP-PGFE)[22] has achieved excellent accuracy on J-TEXT by employing physics-guided feature engineering on the raw diagnostic signals. Although IDP-PGFE has the ability to train with limited data, it performs poorly on smaller datasets and is still difficult to directly apply to future tokamaks. Therefore, we need to explore a cross-tokamak approach to reduce the requirement of disruption predictor on target tokamak data. IDP-PGFE has the following two advantages in cross-tokamak disruption prediction: (1) The required input features for IDP-PGFE are diagnostic-independent and possess physical information. Physics-guided feature extraction (PGFE) processes the raw diagnostic signals from different tokamaks into a unified format, which to some extent aligns certain feature information across tokamaks. Due to PGFE does not rely on training a feature extractor, it no longer requires data to train features when performing feature extraction on the target tokamak. (2) IDP-PGFE has a certain level of interpretability, which can help researchers understand what the model has learned. This can provide researchers with greater confidence in applying the disruption predictor to future tokamaks and may guide them in gaining insights into potential improvements. These natural advantages enable it to have inherent strengths in cross-tokamak disruption prediction tasks.

However, when there are significant differences in size, operational regimes, and even configuration among tokamaks. These issues will be encountered when transferring existing tokamak disruption predictors to future tokamaks. In this work we selected J-TEXT as the existing tokamak and EAST as the future tokamak. J-TEXT is a medium-sized circular section tokamak with a full-carbon wall. The standard J-TEXT discharge can only last 0.7-0.8 seconds. All the discharges are ohmic discharge. In contrast, EAST is a larger-sized elliptical section tokamak with a metal wall. The standard EAST discharge can last 7-8 seconds. The long-pulse discharges can last even tens of seconds. There are also H-mode discharges in EAST. Therefore, compare to the existing tokamak like J-TEXT, EAST could be treated as a future tokamak. Although PGFE can reduce some of the data distribution differences between tokamaks, difference in tokamaks might result in distinct decision boundaries. Consequently, even when applying PGFE to cross-tokamak disruption prediction, it might still be necessary to leverage transfer learning to enhance the effectiveness of cross-tokamak prediction. CORrelation ALignment (CORAL)[36] is a simple, widely-used, and efficient domain adaptation method. CORAL minimizes domain shift by aligning the second-order statistics of source and target distributions, without training or adjustment of any hyperparameters. CORAL aligns the source and target domains in a "frustratingly easy" way, which is lighter and more interpretable. Therefore, in this work, we adopt CORAL as the domain adaptation method for cross-tokamak disruption prediction approach.

In this paper, we present a novel approach for cross-tokamak disruption prediction based on PGFE and CORAL. It represents the first attempt to apply domain adaptation techniques to the task of disruption prediction. To simulate scenarios where significant differences may exist between future and existing tokamaks, J-TEXT is considered as an existing tokamak, while EAST is regarded as a future tokamak. The following section will provide an overview of the cross-tokamak approach, which involves the PGFE application on both J-TEXT and EAST and the adaptation of CORAL to be more suitable for disruption prediction tasks (called supervised CORAL, S-CORAL). Section 3 introduced the dataset used in this work, encompassing J-TEXT and EAST. The cross-tokamak result on EAST is followed in section 4, which shows the improvement of disruption prediction performance by applying S-CORAL. In section 5, we investigate the reasons behind the good performance of S-CORAL in cross-tokamak disruption prediction. Section 6 will briefly discuss the potential of domain adaptation in disruption prediction research and prospects for cross-tokamak disruption prediction. The summary is in section 7.

## 2. The structure of the cross-tokamak disruption prediction

This section will describe the structure of the cross-tokamak disruption prediction based on PGFE and CORAL, which consists of four components, feature extractor, domain adaptation module, disruption classifier and explainer. Compared to IDP-PGFE, there is an additional domain adaptation module used for aligning features. As shown in Figure 1, The first step is pre-processing the raw signal using PGFE to diagnostic-independent and disruptive-related features. Then, CORAL is applied to align the features, mapping the knowledge from the target domain to the source domain. Next, a decision tree-based model called Dropouts meet multiple Additive Regression Trees (DART)[37] will be trained on the mapped data. The trained disruption predictor can be directly used in the target domain. Finally, the trained model can be used for interpretability analysis using SHapley Additive exPlanations (SHAP)[38].

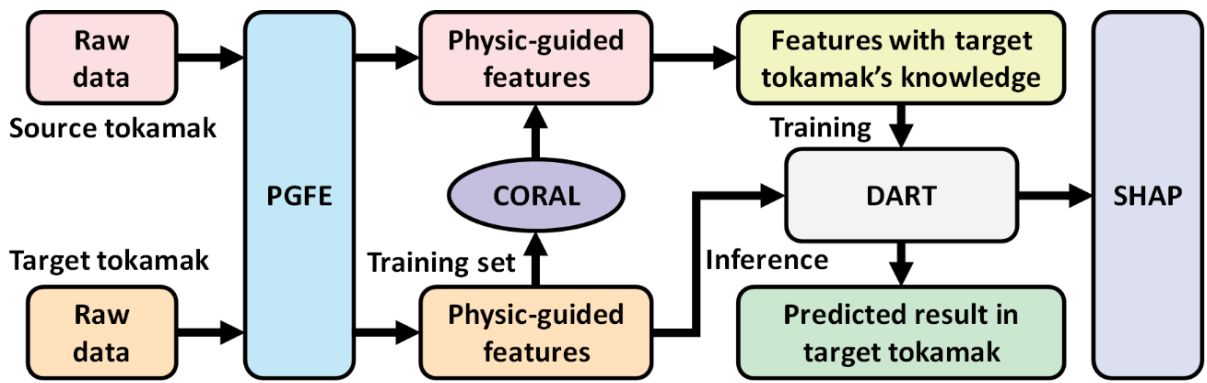

Figure 1 The structure of the cross-tokamak disruption prediction based on PGFE and CORAL. The pink modules represent the source (existing) tokamak, the orange modules represent the target (future) tokamak. The blue module is the feature extractor, PGFE. The purple module is the domain adaptation algorithm, CORAL. The grey module is the classifier, DART. The conch module is the explainer, SHAP.

*2.1 Feature extractor: PGFE on both J-TEXT and EAST*

The diagnostic systems are designed for the engineering and physics requirements of each tokamak. Therefore, it is nearly impossible to directly use diagnostic signals for cross-tokamak disruption prediction. In previous studies on cross-tokamak disruption prediction, different data pre-processing methods[26–28] have been employed to standardize the diagnostic data from different tokamaks into a consistent input format. In this work, PGFE as a feature extractor not only extracts disruptive-related features but also standardizes the input format simultaneously.

The disruption classification for EAST has been analysed[19] and found that impurity radiation, density limit, VDE, and MHD instabilities are also the main causes for EAST disruption. Therefore, the physics-guided features still have been extracted based on MHD instabilities, radiation, density related disruption and basic plasma control system (PCS) signals, just like IDP-PGFE. The diagnostics are different in J-TEXT and EAST; therefore, it is necessary to adjust the parameters in the feature extraction algorithm or redesign the algorithm specifically for EAST. Therefore, PGFE is not an algorithm solely designed for a specific tokamak. Instead, it is necessary to design a unique algorithm for each tokamak to ascertain consistent or analogous physics features. The features for J-TEXT and EAST are listed in Table 1. The numbers followed the name of diagnostic in the third column are the number of channels used for feature extraction. The first number represents the number of channels used in J-TEXT, while the second number represents the number of channels used in EAST.

MHD instabilities related features have been proven to make significant contributions to the disruption prediction task in J-TEXT. However, due to the assumption of circular cross-sections based on J-TEXT, it is difficult to extract *mode_number_m* (*MNM*) feature on EAST. In the future research, if Mirnov data from the entire cross-section of EAST are available, new algorithms can then be designed to compute *MNM* in EAST. Radiation related and density related features are primarily provided by Soft X-ray (SXR)[39,40], Absolute Extreme Ultraviolet (AXUV)[41–43] arrays and polarimeter-interferometer arrays (Far-infrared three-wave polarimeter-interferometer, FIR[44] in J-TEXT and POlarimeter-INTerferometer, POINT[45] in EAST). The cross-tokamak disruption prediction from J-TEXT to EAST demonstrates an advantage of PGFE in handling array signals. Even though the number of SXR, AXUV and polarimeter-interferometer array channels differs between J-TEXT and EAST, PGFE can transform them into the same set of features. Four new features, *n = 1 phase*, *v_loop*, $B_t/I_P$, $I_P\_diff$, are extracted additionally. It is possible that the phase after natural locking may be affected by the inherent differences in the error fields of J-TEXT and EAST. Therefore, *n = 1 phase* has been considered in this work, which could also be calculated like *n = 1 amplitude*[46] by fitting four locked mode detectors. Due to the different wall conditions between EAST and J-TEXT, it is difficult to solely assess the impurity situation only using the *CIII* signal. Therefore, *v_loop* has been introduced to reflect the overall impurity condition. The feature $B_t/I_P$ can approximately represent the information of the boundary safety factor, while $I_P\_diff$ is used to express the variation of the plasma current. As a result, 90 channels of diagnostics in J-TEXT and 65 channels of diagnostics in EAST have been extracted into 25 features. The extraction algorithms and significance of most features have been extensively discussed in this research[22]. Therefore, in this paper, we will provide a brief introduction to them.

Table 1 Descriptions and symbols of all the features for J-TEXT and EAST

| Types of features | Relation to disruption | Channels of diagnostics (J-TEXT/EAST) | Symbol |
|---|---|---|---|
| MHD instabilities related | 2/1 magnetic island growth; Multi-magnetic island overlaps; Locked mode | Mirnov probe in poloidal array<br>Mirnov probes in toroidal array (2/2)<br>Locked mode detectors (4/4) | Mir_abs<br>Mir_fre<br>Mir_Vpp<br>mode_number_n (MNN)<br>n=1 amplitude<br>n = 1 phase |
| Radiation related | Temperature hollowing; Edge cooling; | Soft x-ray (SXR) array (30/20)<br>Central channel of SXR array<br>Central channel of CIII radiation array<br>Absolute eXtended Ultra Violet (AXUV) array (30/21)<br>Loop voltage | $SXR_{kurt\ (skew,\ var)}$<br>$SXR_{core}$<br>CIII<br>$AXUV_{kurt\ (skew,\ var)}$<br>v_loop |
| Density related | Density limit | polarimeter-interferometer array (17/11)<br>Central channel of polarimeter-interferometer array | $DEN_{kurt\ (skew,\ var)}$<br>$n_{e0}$ |
| Basic PCS signals | Plasma out of control; | Toroidal field and plasma current<br>Horizontal and vertical displacements,<br>Ratio of toroidal field to plasma current<br>Plasma current variation rate | $B_t$, $I_p$<br>$d_r$, $d_z$<br>$B_t/I_P$<br>$I_p\_diff$ |

The feature extraction algorithm for Mirnov probes mainly relies on the Fast Fourier Transform (FFT). $X^i(f)$ represents the Fourier transform of the $i^{th}$ time window (slice). "*Mir_abs*" and "*Mir_fre*" are the intensity and frequency of $X^i(f)$ with the highest spectral intensity. Here, we also did not use the integral Mirnov signals to avoid the zero-drift uncertainty between different tokamaks and discharges that contributed to our prediction. $X_1^i(f)$ and $X_2^i(f)$ represent the Fourier transform of two Mirnov probe signals of each slice. The low-pass filter with a cut-off frequency of 50 kHz and 10 kHz are designed in Mirnov data of J-TEXT and EAST before FFT, respectively. Their cross spectral density (CSD) can be expressed as

$$P_{12}^j(f) = X_1^{j*}(f) X_2^{j*}(f) = A(f) e^{i \delta_{12}^j(f)} \qquad (2\text{-}1)$$

where $P_{12}^i(f)$ is the cross spectral density between two Mirnov probes, $A(f)$ is the absolute value of $P_{12}^i(f)$, and $\delta_{12}^i(f)$ is the phase of $P_{12}^i(f)$. For the two toroidal Mirnov probes, the mode number $n$ can be calculated as

$$n = \frac{\delta_{12}^j(f)}{\varphi} \qquad (2\text{-}2)$$

where $\varphi$ is the toroidal separation between the two Mirnov probes. The toroidal separation is 22.5° and 45° in J-TEXT and EAST, respectively. Only the frequency component with a coherence larger than 0.95 will be considered. The magnetic field measured by the locked mode (LM) detector can be expressed as

$$B_r(\theta, \varphi) = \sum b_r^{n=i} \cos(m\theta + n\varphi + \xi^{n=i}) \qquad (2\text{-}3)$$

where $\theta$ and $\varphi$ are the poloidal and toroidal location of the LM detector, $m$ and $n$ are the poloidal and toroidal mode number, and $\xi$ is the spiral phase. If only consider the main component in J-TEXT and EAST, which are $n = 0$, 1 and 2, for the detector on the middle plane $\theta = 0$ (The following are all based on this situation), equation (2-3) can be expressed as

$$B_r(\theta = 0, \varphi) = b_r^{n=0} + b_r^{n=1} \cos(\varphi + \xi^{n=1}) + b_r^{n=2} \cos(2\varphi + \xi^{n=2}) \qquad (2\text{-}4)$$

$n = 1$ amplitude, $b_r^{n=1}$ can be calculated through two LM detectors with $\Delta\varphi = \pi$, which is shown in equation (2-5)

$$b_r^{n=1} \cos(\varphi + \xi^{n=1}) = \frac{B_r(\varphi) - B_r(\varphi + \pi)}{2}. \qquad (2\text{-}5)$$

Then $b_r^{n=1}$ ($n = 1$ amplitude) and $\xi^{n=1}$ ($n = 1$ phase) can be calculated by fitting two pairs of these LM detectors.

In the previous work, we calculated the higher-order statistics (HOS) of the 1D signals from the SXR, AXUV and polarimeter-interferometer arrays for radiation and density related features in J-TEXT. This is a compromise solution that aims to incorporate 1D profile information into the model while avoiding the impact of inversion errors on the prediction results. In cross- machine disruption prediction, this method can also unify array information from different tokamaks with varying numbers of channels into a single feature input model, greatly enhancing the flexibility of the model. The variance (*var*), skewness (*skew*), and kurtosis (*kurt*) of the array signals have been selected to extract the 1D signals to 0D features.

### 2.2 Domain adaptation module: CORAL

The goal of domain adaptation is to bridge the gap between the source and target domains by transferring knowledge learned from the source domain to the target domain. This transfer of knowledge enables the model to generalize well and make accurate predictions on the target domain despite the differences. CORAL[36] is an efficient, lighter and more interpretable domain adaptation method, which minimizes domain shift by aligning the second-order statistics of source and target distributions. CORAL aligns the distributions by re-colouring the whitened source features using the covariance of the target distribution. It is a simple and more interpretable method that involves two main computations: (1) calculating covariance statistics in each domain and (2) applying the whitening and re-colouring linear transformation to the source features. Afterward, supervised learning can proceed as usual by training a classifier on the transformed source features. The brief mathematical derivation and assumptions underlying the implementation of CORAL is followed.

Supposing that the source domain data is $D_S = \{\vec{x}_i\}, \vec{x}_i \in \mathbb{R}^D$ and target domain data is $D_T = \{\vec{u}_i\}, \vec{u}_i \in \mathbb{R}^D$, Here $\vec{x}_i$ and $\vec{u}_i$ are the $D$-dimensional input feature representations. A linear transformation $A$ has been applied to the original source features and the Frobenius norm has been used as the matrix distance metric. As a result, the distance between the second-order statistics (covariance) of the source and target features could be minimized.

$$\min_A \|C_{S'} - C_T\|_F^2 = \min_A \|A^T C_s A - C_T\|_F^2 \qquad (2\text{-}6)$$

where $C_{S'}$ is covariance of the transformed source features $D_S A$. and $\|\cdot\|_F^2$ denotes the matrix Frobenius norm. The optimal solution of $A$ can be expressed by deduction as:

$$A^* = (U_S \Sigma_S^{+\frac{1}{2}} U_S^T)(U_{T[1:r]} \Sigma_{T[1:r]}^{\frac{1}{2}} U_{T[1:r]}^T) \qquad (2\text{-}7)$$

Where $r = \min(r_{C_S}, r_{C_T})$, $r_{C_S}$ and $r_{C_T}$ denote the rank of $C_S$ and $C_T$, respectively. Since $C_S$ and $C_T$ are symmetric matrices, conducting singular value decomposition (SVD) on $C_S$ and $C_T$ gives $C_S = U_S \Sigma_S U_S^T$ and $C_T = U_T \Sigma_T U_T^T$, respectively. $\Sigma^+$ is the Moore-Penrose pseudoinverse of $\Sigma$. The final algorithm can be written in four lines of MATLAB[36] and Python code as illustrated in Algorithm 1.

**Algorithm 1** CORAL for MATLAB and Python code

| |
|---|
| **Input:** Source Data $D_S \in \mathbb{R}^{n_s \times n_{features}}$, Target Data $D_T \in \mathbb{R}^{n_T \times n_{features}}$ |
| **Output:** Adjusted Source Data $D^*_S$ |
| **MATLAB code:** <br> $C_S$ = cov ($D_S$) + eye (size ($D_S$,2)) <br> $C_T$ = cov ($D_T$) + eye (size ($D_T$,2)) <br> $D_S$ = $D_S$ * $C_S$ ^ (-0.5) % whitening source <br> $D^*_S$ = $D_S$ * $C_T$ ^ (0.5) % re-colouring with target covariance |
| **Python code:** <br> $C_S$ = np.cov ($D_S$.T) + np.eye ($D_S$.shape[1]) <br> $C_T$ = np.cov ($D_T$.T) + np.eye ($D_T$.shape[1]) <br> $A_{CORAL}$ = np.dot (scipy.linalg.fractional_matrix_power($C_S$, -0.5), <br> scipy.linalg.fractional_matrix_power($C_T$, 0.5)) <br> $D^*_S$ = np.real(np.dot($D_S$, $A_{CORAL}$)) |

CORAL was originally designed as an unsupervised domain adaptation (UDA) algorithm[31]; however, the target data are actually labelled in cross-tokamak disruption prediction. Therefore, we improved CORAL to a supervised domain adaptation (SDA) version and applied it in this paper. The idea for improvement is similar to the approach of designing SDA in machine learning research[47], which is to consider label information when CORAL minimizes the distance between the covariance matrices in different domains. We called the SDA version as supervised CORAL (S-CORAL) and the UDA version as unsupervised CORAL (U-CORAL) in this paper. The flowchart of S-CORAL and U-CORAL is shown in Figure 2. U-CORAL colours the covariance of the whole target data to the source data, thus it cannot effectively consider the information from labelled data. This is a waste for disruption prediction tasks with clearly defined positive and negative samples. S-CORAL can effectively consider the information from labelled data by aligning the disruptive samples and non-disruptive samples separately between the target and source domains.

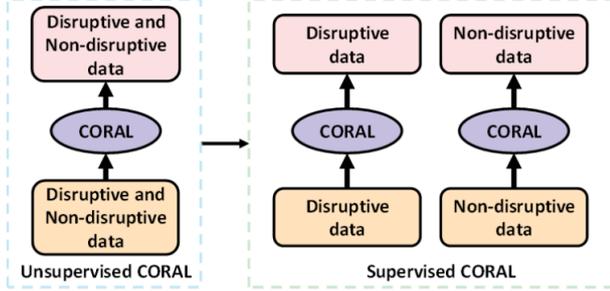

Figure 2 The flowchart of S-CORAL and U-CORAL. The colours of the modules are consistent with Figure 1.

### 2.3 Disruption classifier: DART and Explainer: SHAP

IDP-PGFE has proved that DART[37] is a suitable disruption classifier, which could realize high-performance disruption prediction on J-TEXT. DART is a Gradient boosting decision tree (GBDT)[48] based algorithm, which uses the trick of the dropout setting in deep neural networks[37] to drop the existing decision trees randomly. DART iteratively optimizes the boosted trees from the remaining set of decision trees could alleviate the over-sensitive problem on the contributions of the few initially added trees.

IDP-PGFE has also proved that SHAP[38] is a suitable model explainer, which could realize enough interpretability on J-TEXT. SHAP is based on the game theoretically optimal Shapley values [49], which is a method from coalitional game theory to figure out how fairly distribute the "pay-out" (prediction) among the "players" (features). The Shapley value is the contribution of a feature value to the difference between the actual prediction and the mean prediction when given the current set of feature values. SHAP is also an additive feature attribution method, which uses a simpler explanation model than any interpretable approximation of the original model. Explanation models use simplified inputs $x'$ that map to the original inputs through a mapping function $x = h_x(x')$. An explanation model $g$ can be expressed as:

$$g(z') = \phi_0 + \sum_{j=1}^{M} \phi_j z_j' \qquad (2\text{-}10)$$

Where $z' \in \{0,1\}^M$, $z' = 1$ means that the corresponding feature value is "present" and 0 that it is "absent". $M$ is the number of simplified input features, and $\phi_i \in R$ is the feature attribution for a feature $j$, the Shapley values. The Shapley values $\phi(f, x)$ expressed as:

$$\phi_j(f,x) = \sum_{z' \subseteq x'} \frac{|z'|!(M-|z'|-1)!}{M!}[f_x(z') - f_x(z' \setminus j)] \qquad (2\text{-}11)$$

where $|z'|$ is the number of non-zero entries in $z'$ and $z' \setminus j$ denotes setting $z' = 0$. The Shapley values $\phi_i(f, x)$, explaining a prediction $f(x)$, are an allocation of credit among the features in $x$ (the features extracted through PGFE) and are the only allocation satisfying three desirable properties. The first one is local accuracy, which ensure the explanation model $g$ at least match the predictor $f$. The second one is missingness to ensure features missing in the original input to have no impact, which means if $z' = 0$, the importance attributed is also 0. The third one is consistency (also called monotonicity in game theory researches), which states that if a feature is more important in one model than another, the importance attributed to that feature should also be higher.

## 3. Dataset description and training approaches

### 3.1 Dataset description

The introduction provides a brief overview of the differences between J-TEXT and EAST. Compared to the "existing tokamak" J-TEXT, EAST can be treated as a "future tokamak". This section will introduce the two tokamaks in detail first, then describe the dataset selected to train, valid and test.

J-TEXT is a medium-sized tokamak with a major radius $R = 1.05$m and a minor radius $a = 0.25$m[50]. J-TEXT is equipped with a comprehensive diagnostic system comprising over 300 channels of various diagnostics. In the limiter configuration, typical discharges on the J-TEXT are characterized by a plasma current ($I_P$) of approximately 200 kA, a toroidal field ($B_t$) of around 2.0 T, a pulse length of 700 - 800 ms, plasma densities ($n_e$) ranging from 1 to $7 \times 10^{19}$ m$^{-3}$, and an electron temperature ($T_e$) of about 1 keV. The typical resistive time scales in J-TETX is about 25ms ($\tau_R \approx 25$ ms). EAST is an ITER-like fully super-conducting tokamak with a major radius $R = 1.85$m and a minor radius $a = 0.45$m[51]. It shares common diagnostic systems with J-TEXT, including measurements of radiation, displacement, locked modes, MHD instabilities, plasma current, plasma density, and other diagnostics. In the divertor configuration, typical discharges on the EAST are characterized by a plasma current ($I_P$) of approximately 450 kA, a toroidal field ($B_t$) of around 1.5 T, a pulse length of approximately 10 s, and a $β_N$ of around 2.1. The typical resistive time scales in EAST is larger than 500ms ($τ_R \geqslant 500$ ms).

The dataset and its split are similar to IDP-PGFE[22]. The dataset of J-TEXT contains 1734 (378 disruptive) discharges out of 2017-2018 campaigns with the accessibility and consistency of the diagnostics channels. All types of disruptions were included except intentional ones triggered by massive gas injection (MGI) or shattered pellet injection (SPI) and engineering tests. The training, validation and test sets are selected randomly from the 2017-2018 campaigns. A split of datasets is shown in Table 2. 1354 discharges (188 disruptive) are selected as the training set. 160 discharges (80 disruptive) are selected as the validation set. 220 discharges (110 disruptive) are selected as the test set. As for EAST, the dataset and its split are similar to our previous work[35]. The training, validation and test sets are still selected randomly. A total of 1896 discharges (355 disruptive) discharges are selected as the training set and 120 discharges (60 disruptive) are selected as the validation set for the full EAST dataset model. 110 (10 disruptive) discharges are selected as training set for the cross-tokamak models. 360 discharges (180 disruptive) are selected as the test set full EAST dataset model and cross-tokamak models to make a fair comparison.

Table 2 Split of datasets of the predictor

|  | J-TEXT | EAST (full data) | EAST (cross-tokamak) |
| --- | --- | --- | --- |
| Training | 1354 (188) | 1896 (355) | 110 (10) |
| Validation | 160 (80) | 120 (60) | / |
| Test | 220 (110) | 360 (180) | |

All discharges are split into slices from the flat-top of plasma current to current quench (CQ) time per 1ms, the same as the sampling rate of plasma current in EAST. An automatic criterion has been applied to detect sudden large drops in $I_P$, and marked the beginning of the drop as CQ time with 1 ms resolution. Then the results are visually checked and corrected by human experts. The phase between the CQ time and a time threshold indicates the unstable phase of each disruptive discharge. The unstable phase and time threshold of each discharge can be determined manually [52] by a statistical analysis, either equal for each discharge [12] or individually for each discharge [11,53]. An automatic approach has been used to determine the unstable phase and time threshold by finding the best performance of the model by scanning the time threshold from 5ms to 50ms in J-TEXT and 5ms to 500ms in EAST. The time threshold equalled to 25ms for J-TEXT and 125ms for EAST before CQ time achieved the best performance. The "unstable" samples in disruptive charges are labelled as "disruptive", and all the samples in non-disruptive discharges are labelled as "non-disruptive". This sample partitioning is primarily based on two considerations: a) The number of non-disruptive discharges is significantly higher than the number of disruptive discharges, hence there is already an ample amount of non-disruptive samples available. b) This fixed labelling approach for each shot introduces some erroneous prior information. To minimize the introduction of such information, non-disruptive samples from disruptive discharges are not used. Although, the non-disruptive samples are not considered from disruptive discharges, the significant imbalance of the dataset is still existed. Therefore, we increased the weights of disruptive samples and randomly dropped a portion of non-disruptive samples to balance the two kinds of samples.

### 3.2 The training of the disruption prediction models

In this paper, five models have been trained. The first two models are self-tokamak disruption prediction models, distinct from the cross-tokamak disruption prediction model. The J-TEXT model demonstrates performance in the source domain, serving as the base for the cross-tokamak models. The EAST model, trained using the full data training set of EAST as shown in Table 2, representing the peak of performance. These two models adopt the structure of IDP-PGFE[22] instead of the structure in Figure 1. The third model is the mixing data model, which mixed the training set of J-TEXT and cross-tokamak training set of EAST in Table 2. The mixing data model also used the structure of IDP-PGFE. The last two models are CORAL models, which used the structure in Figure 1. After applying PGFE to the entire dataset, the cross-tokamak training set from EAST is aligned with the J-TEXT training set in Table 2 by CORAL. For validation, we exclusively use the J-TEXT validation set. Once training is completed, we will use the EAST test set to make predictions. Finally, a thorough analysis using SHAP is conducted to find out what have the models learned and where could be improved for the cross-tokamak disruption prediction.

## 4. Predictive performances of the models

This section shows the predictive performances of various models. The self-tokamak models of J-TEXT and EAST will be first shown as the benchmark model for the cross-tokamak in section 4.1. Then section 4.2 will compare the mixing data model, unsupervised model and supervised model for cross-tokamak disruption. The hyperparameter search determines the hyperparameters of each best performance model in this section.

Disruption prediction is a binary classification task, where the performance is often evaluated using a confusion matrix. In the context of disruption prediction, True Positive (TP) refers to successfully predicting a disruptive discharge. False Positive (FP) refers to a non-disruptive discharge being incorrectly predicted as disruptive, also known as a false alarm. True Negative (TN) refers to a correctly predicted non-disruptive discharge. False Negative (FN) refers to a disruptive discharge not being predicted as disruptive. Both missed alarms and delayed alarms are considered as FN. It is important to note that a short warning time should be considered as a delayed alarm, considering the requirements of the Disruption Mitigation System (DMS). For J-TEXT, any predicted disruption with a warning time of less than 10ms is considered FN. For EAST, any predicted disruption with a warning time of less than 30ms is considered FN. The evaluation indicators of a disruption predictor are the receiver operating characteristic curve (ROC), which included true positive rate (TPR), false positive rate (FPR) and area under the ROC curve (AUC). TPR and FPR are calculated as follows:

$$\text{TPR} = \frac{TP}{TP+FN} \quad (4\text{-}1)$$

$$\text{FPR} = \frac{FP}{FP+TN} \quad (4\text{-}2)$$

The DART will give a result between "0" ("non-disruptive") and "1" ("disruptive"), which can be binarily classified by manually setting a model threshold. Each model threshold corresponds to a set of TPR and FPR. The ROC curve is created by plotting TPR against FPR at various threshold settings.

The normalization process is performed independently for both tokamaks. It is worth noting that, in the normalization process of cross-tokamak disruption prediction on EAST, only 110 known discharges were used to calculate the normalization parameters, aiming to simulate real application scenarios. The inputs are normalized with the *z-score* method, which $x_{norm} = \frac{x - mean(x)}{std(x)}$.

## 4.1 The self-tokamak models of J-TEXT and EAST based on PGFE

In this part, the self-tokamak model of J-TEXT and EAST was trained with full training set data and the ROC curves are shown in Figure 3. The yellow and orange line represents the ROC curves of the J-TEXT model and the EAST model, respectively. The AUC value of J-TEXT model is 0.971 and the AUC value of EAST model is 0.936. The predict performance of the J-TEXT model is not as good as the pervious IDP-PGFE model, due to the lack of some features. To maintain consistency with the sampling rate of EAST, the J-TEXT model takes each sample as 1ms, while the pervious IDP-PGFE model takes each sample as 0.1ms, which may also impact the model performance.

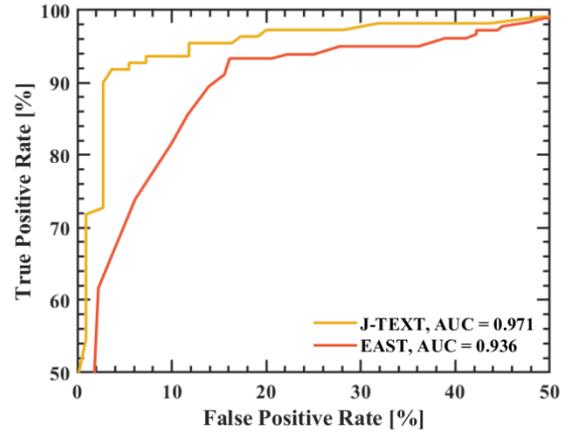

Figure 3 The ROC curves of the self-tokamak model of J-TEXT and EAST. The FPR axis is from 0% to 50%. The TPR axis is from 50% to 100%. The yellow and orange line represents the ROC curves of J-TEXT model and EAST model, respectively.

Figure 4 shows the accumulated percentage of disruption predicted versus warning time with the model threshold = 0.56 for J-TEXT model and the model threshold = 0.93 for EAST model. Due to J-TEXT being a small-sized tokamak with a relatively smaller time scale for disruption to take place, the warning time is shorter than that for JET, EAST, DIII-D, or other large and medium sized tokamaks. Therefore, the warning time could be selected as 10ms. The electromagnetic particle injector (EPI) could react by the trigger advanced 10ms 54. The warning time is selected as 30ms for EAST. For J-TEXT model, the warning time of 30ms should also ensure a considerable accumulated percentage (TPR>90%) of disruption is predicted for other mitigation methods to react. For EAST model, the warning time of 50ms should also ensure a considerable accumulated percentage (TPR>90%) of disruption is predicted for other mitigation methods to react. Therefore, the final performance of J-TEXT and EAST self-tokamak model is TPR = 93.64%, FPR = 8.18% with a tolerance of 10ms and TPR = 93.33%, FPR = 16.11% with a tolerance of 30ms, respectively. The average and median of the warning time for the J-TEXT model are both 0.14 seconds, while for the EAST model, the average warning time is 1.34 seconds and the median warning time is 0.75 seconds. Due to the presence of long pulse discharges in EAST, the average warning time may be significantly influenced. Therefore, the median value better reflects the overall warning time of the EAST model.

As a result, PGFE has successfully achieved good performance for both J-TEXT and EAST self-tokamak models by considering the common physics-guided features shared by J-TEXT and EAST tokamaks. This provides a base for the next cross-tokamak works, serving as a model and reference.

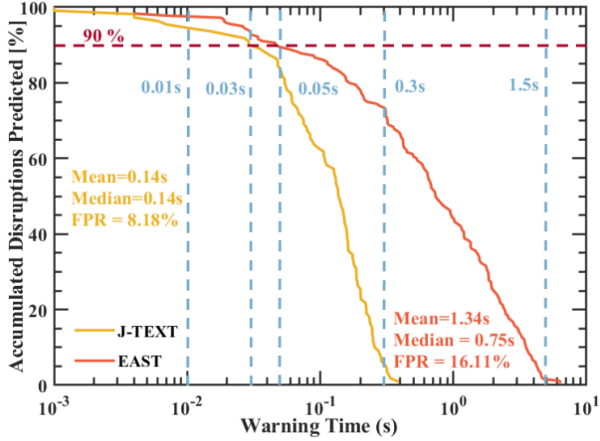

Figure 4 The accumulated percentage of disruption predicted versus warning time. The model threshold is 0.56 for J-TEXT model (yellow) and the model threshold is 0.93 for EAST model (orange). The red dashed line represents the accumulated percentage of disruption predicted equals to 90%. The light blue dashed lines represent the warning time of 0.01s (10ms), 0.03s (30ms), 0.05s (50ms), 0.3s (300ms) and 1.5s (1500ms).

### 4.2 Models of cross-tokamak disruption prediction from J-TEXT to EAST based on PGFE and CORAL

In this part, three models of cross-tokamak disruption prediction from J-TEXT to EAST, whose strategies are mixing data, U-CORAL and S-CORAL, are described and analysed. Although IDP-PGFE has the ability to train with limited data, it performs poorly on smaller datasets, such as 10 disruptive discharges and 120 non-disruptive discharges in J-TEXT. It is worth noting that even when training with limited data from a single tokamak, it is still necessary to have a certain amount of validation data to prevent overfitting and select the best model. Therefore, the functionality of training with limited data is not suitable for scenarios such as 10 disruptive discharges and 100 non-disruptive discharges in EAST.

The models of five cases, including the three cross-tokamak models, are shown in Table 3. In the "EAST data" column, the numbers represent the "total number of discharges (number of disruptive discharges) ". The model of case 1 is a baseline model for the cross-tokamak disruption prediction, which is directly testing EAST data on the J-TEXT self-tokamak model called zero-shot test. The model of case 2 adopts the strategy of mixing limited EAST data with J-TEXT data to train a model. This strategy is also commonly used in cross-tokamak disruption prediction approaches. The model of case 3 directly uses CORAL to map the knowledge from EAST data into J-TEXT data for training. We called the data strategy of this case is CORAL and the CORAL strategy is unsupervised. The model of case 4 uses S-CORAL to map the knowledge from both disruptive and non-disruptive EAST data to J-TEXT data separately, and then trains the model. We also called the data strategy of this case is CORAL and the CORAL strategy is supervised. The model of case 5 is the self-tokamak model of EAST as a benchmark, which means the peak performance of the datasets and has been described in section 4.1.

Table 3 Five models of cross-tokamak disruption prediction

| Case No. | EAST Data | Data Strategy | CORAL Strategy | AUC |
|---|---|---|---|---|
| 1 | None | / | / | 0.642 |
| 2 | 110 (10) | Mixing | / | 0.764 |
| 3 | 110 (10) | CORAL | Unsupervised | 0.797 |
| 4 | 110 (10) | CORAL | Supervised | **0.890** |
| 5 | 1896 (355) | Full data | / | 0.936 |

The ROC curve of these five models are shown in Figure 5. The light-blue line represents the ROC curves of the J-TEXT benchmark model by zero-shot test with the AUC value of 0.642. The yellow line represents the ROC curves of the mixing data model with the AUC value of 0.764. The green line represents the ROC curves of the U-CORAL model with the AUC value of 0.797. The navy-blue line represents the ROC curves of the S-CORAL model with the AUC value of 0.890. The orange line still represents the EAST model with the AUC value of 0.936. Similar to previous studies on cross-tokamak disruption prediction, the mixing data model does improve the prediction performance compared to direct zero-shot testing (AUC value from 0.642 to 0.764). However, for cross-tokamak disruption prediction with significant differences in device and discharge parameters, such as from J-TEXT to EAST, the performance of the mixing data model is still unacceptable. The U-CORAL model shows a little improvement in prediction performance compared to the mixing data model, but the improvement is not significant (AUC value from 0.764 to 0.797). However, the S-CORAL could significantly improve the performance compared to other strategies (AUC value from 0.797 to 0.890) and has a smaller performance gap compared to the EAST model.

Figure 6 shows the accumulated percentage of disruption predicted versus warning time with the model threshold = 0.93 for EAST model, the model threshold = 0.01 for mixing data model, the model threshold = 0.31 for U-CORAL model and the model threshold = 0.58 for S-CORAL. The warning time is selected as 30ms for EAST test set. Our principle for selecting the model threshold is to first ensure the TPR under this threshold higher than 90%. Based on this criterion, we further choose a model threshold that achieves a lower FPR. However, for the mixing data model, the highest TPR is 73.33% under the model threshold is 0.01. Therefore, the final performance of mixing data model, U-CORAL model and S-CORAL model is TPR = 73.33%, FPR = 27.78%, TPR = 90.56%, FPR = 46.67% and TPR = 90%, FPR = 25.56% with the tolerance of 30ms, respectively. The average warning time is 1.48 seconds and the median warning time is 0.72 seconds for mixing data model. The average warning time is 1.85 seconds and the median warning time is 1.54 seconds for U-CORAL model. The average warning time is 1.43 seconds and the median warning time is 0.74 seconds for S-CORAL model. Except for the U-CORAL model, the average and median warning times of the other two models are similar to the EAST model.

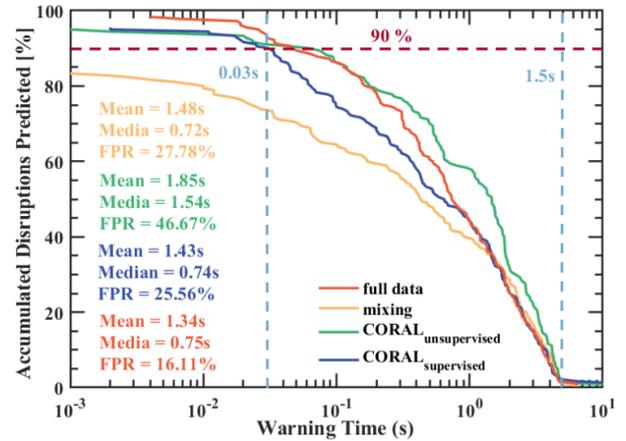

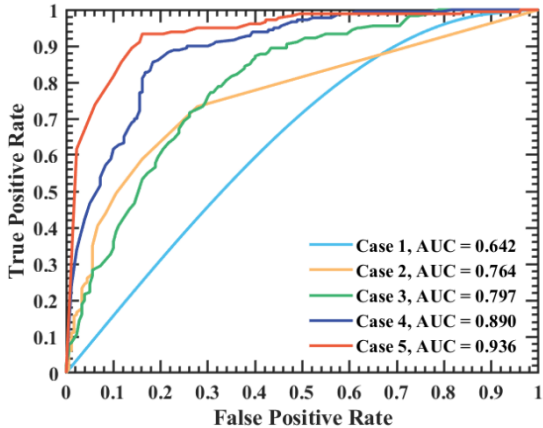

Figure 5 The ROC curves of the models of five cases. Five coloured lines represent the models of five cases, respectively (case 1 – light-blue, case 2 – yellow, case 3 – green, case 4 – navy-blue and case 5 – orange).

It can be concluded that the application of CORAL, outperforms the previously widely used method of mixing data in cross-tokamak disruption prediction. S-CORAL model further improves the performance of cross-tokamak disruption prediction, achieving the TPR of 90%, FPR of 25.56%, and AUC value of 0.89. This performance is close to that of the EAST self-tokamak model trained with full data. Therefore, in terms of performance, cross-tokamak disruption prediction based on PGFE and CORAL is a competitive approach for achieving cross-tokamak disruption prediction in future tokamaks.

Figure 6 The accumulated percentage of disruption predicted versus warning time. The model threshold is 0.93 for EAST self-tokamak model (orange), the model is 0.01 for mixing data model (yellow), the model threshold is 0.31 for U-CORAL model (green) and the model threshold is 0.58 for S-CORAL model (navy-blue). The red dashed line represents the accumulated percentage of disruption predicted equals to 90%. The light blue dashed lines represent the warning time of 0.03s (30ms) and 1.5s (1500ms).

## 5. Interpretability study of the cross-tokamak disruption prediction based on PGFE and CORAL

This section will describe the interpretability study. The objective of the interpretability study is to investigate why the S-CORAL model can outperform the mixing data model and U-CORAL. It can also provide insights and valuable experience for future applications on other future tokamaks such as ITER and SPARC. Section 5.1 will investigate how S-CORAL aligned training data distribution between J-TEXT and EAST, which can be called intrinsic interpretability. Section 5.2 will use SHAP to explore the differences in knowledge learned by the mixing data, U-CORAL, and S-CORAL models compared to the knowledge learned by the full data trained EAST self-tokamak model on the test set. This interpretable approach can be called post-hoc interpretability. A method to evaluate this difference has been identified to demonstrate that S-CORAL indeed learns knowledge closer to that of the EAST self-tokamak model.

### 5.1 Data distribution analysis

Although PGFE can align the diagnostic signals of EAST and J-TEXT into physics-guided features, the large gap between the two tokamaks means that the decision boundary for each tokamak will differ. For instance, the same physics phenomena or parameter changes might trigger a disruption warning in J-TEXT, but not necessarily cause a disruption in EAST. The purpose of normalization and CORAL is to align each feature as much as possible in terms of data distribution. S-CORAL could perform better than U-CORAL and mixing data because more features could be aligned better than in the other two cases. The probability density of four typical normalized features $n_{e0}$, $n = phase$, $SXR\_array\_skew$ and $SXR\_array\_kurt$ have shown in Figure 7 and Figure 8. Figure 7 represents the probability density for non-disruptive data, while Figure 8 represents the probability density for disruptive data. The yellow region and lines represent the probability density of J-TEXT data. The orange region and lines represent the probability density of EAST data. The green region and lines represent the probability density of U-CORAL data. The navy-blue region and lines represent the probability density of S-CORAL data.

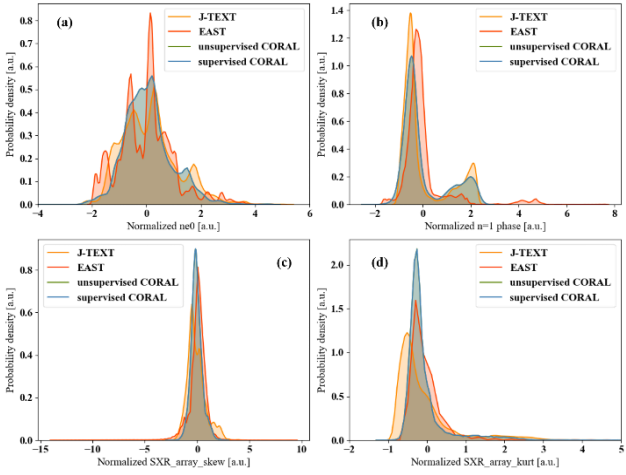

Figure 7 The probability density for non-disruptive data of four typical normalized features (a) $n_{e0}$, (b) $n = 1\ phase$, (c) $SXR\_array\_skew$ and (d) $SXR\_array\_kurt$.

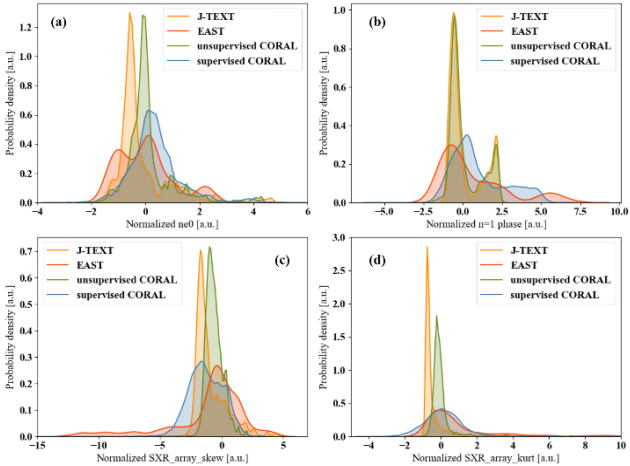

Figure 8 The probability density for disruptive data of four typical normalized features (a) $n_{e0}$, (b) $n = phase$, (c) $SXR\_array\_skew$ and (d) $SXR\_array\_kurt$.

Figure 7 demonstrates that the non-disruptive data distribution between J-TEXT and EAST is already quite similar. This indicates that the process through PGFE and normalization is sufficient to align the distributions of non-disruptive data between J-TEXT and EAST effectively. The non-disruptive data distribution of U-CORAL and S-CORAL is the same, which indicates that the two methods of CORAL will not affect the non-disruptive data distribution. In comparison, Figure 8 shows that the disruptive data distribution of S-CORAL is more similar to the disruptive data distribution of EAST than that of U-CORAL. At the same time, the disruptive data distribution of U-CORAL is more similar to the disruptive data distribution of J-TEXT than that of S-CORAL. After alignment using S-CORAL, the disruptive data distribution contains more information from the distribution of EAST data. But not all features require the S-CORAL for aligning the disruption data distribution, and not all features can be effectively aligned through PGFE and CORAL. The application of PGFE and normalization could align the non-disruptive data distribution of 56% (14/25) features and could align the disruptive data distribution of 36% (9/25) features. The application of S-CORAL could align the disruptive data distribution of 68% (17/25) features.

In summary, the non-disruptive data distribution of over half features could be aligned sufficient by PGFE and normalization, while aligning the disruptive data distribution of most features requires the additional use of S-CORAL.

### 5.2 SHAP analysis

The data distribution analysis shows that the feature-based method PGFE and CORAL could align the data distribution between J-TEXT and EAST. It is a kind of intrinsic interpretability [54] in the interpretable machine learning due to the PGFE and CORAL are kind of rule-based models. SHAP is an attribution-based interpretable approach, which is a kind of post-hoc interpretability.

SHAP provides global interpretability for the models, analysing the contribution of feature variations to the model's output. In this section, the full data trained EAST self-tokamak model is a benchmark model. The three cross-tokamak models (mixing data, U-CORAL and S-CORAL) will be compared to this reference model to analyse which cross-tokamak model has learned more from the EAST data by the similarity of the global interpretability. To make a fair comparison, the dataset for the SHAP is selected as the test set. Figure 9 (b) shows the global SHAP value of different features and their relations with feature value of EAST self-tokamak model. The SHAP results on the test set can be understood as the knowledge the model learned that is applied when distinguishing "disruptive" or "non-disruptive" in the test set. The order of the features represents the contributions of features. The colormap represents the feature value of each feature, red means high and blue means low. The advantage of SHAP in global interpretability is that it cannot only provide the ranking of feature contributions to the model but also indicate whether the variations of the features contributes positively or negatively to the model's predictions. When analysing whether multiple models have learned similar knowledge, the positive or negative contribution of feature variations to the model is more important than the ranking of the feature's contribution. When predicting disruptions, even for physicists, judgments about disruptions might be made based on various features. Different physicists might have varying rankings of feature importance when predicting disruptions. However, regardless of how feature importance is ranked, the variations of the feature contributed to " disruptive " or "non- disruptive " should remain the same. For instance, before a density limit disruption, not only is the density a critical feature, but MARFE, MHD instabilities are also crucial indicators of density limit disruption. No matter the feature importance ranking, the higher the density, the more significant its contribution to the disruption. The distribution of disruption types and causes in the dataset can also affect the ranking of feature contributions. Therefore, the variations of the feature contributed to " disruptive " or "non- disruptive " should be more important than the ranking of the feature contributions.

An evaluation method has been designed to assess whether the knowledge learned by models is similar based on the global interpretability results of SHAP. The core logic of this evaluation method is the counting of features with similar change patterns. Since the full data trained EAST self-tokamak model was trained using a larger amount of EAST data and performed the best, the learned knowledge about EAST disruptions is more comprehensive and accurate. Therefore, the three cross-tokamak models will be compared using the full data trained EAST model as the benchmark. If the variations of any feature in the cross-tokamak model contributed to "disruptive" or "non- disruptive " is the same as that in the benchmark model, then that feature scores positive one point. Such as, the greater the value of the feature $v\_loop$, the higher its contribution to the disruption in the self-tokamak model. The trend of the feature contribution is also consistent in the S-CORAL model. Thus, for the S-CORAL model, the feature $v\_loop$ scores a positive one point. On the contrary, if the variations of any feature in the cross-tokamak model contributed to "disruption" or "non-disruption" is not the same as that in the self-tokamak model, then that feature scores negative one point. For example, in the self-tokamak model, the larger the value of the feature $d_Z$, the higher its contribution to the "disruptive". In the S-CORAL model, the value of $d_Z$ does not contribute to either "disruptive" or "non- disruptive ", which is also not the same as it in the benchmark model. Therefore, for the S-CORAL model, the feature $d_Z$ is given a negative one point. Then, we added the score of each feature to evaluate which model is more similar to the benchmark model.

Figure 9 (a) also shows all the feature scores of three cross-tokamak models by the similarity evaluation with the self-tokamak model. Columns 1,2,3 of the table show the scores of models mixing data, U-CORAL, S-CORAL on each feature, respectively. Except for the first and last rows, each row corresponds to each feature in Figure 9 (b). The last row of the table with the red background colour shows the total score. The rank of total score indicates that the S-CORAL model (scores 7) is the most similar to the self-tokamak model. While, the mixing data model scores -3 and U-CORAL model scores 3 The skewness of the array signals is a measure of the asymmetry of the distribution of the array signals about their mean[22], which can be approximately regarded as reflecting the displacement extent of the plasma measured by the diagnostics. The score of $d_r$, $AXUV_{skew}$, $SXR_{skew}$, and $DEN_{skew}$ in all three cross-tokamak models are -1. It reflects that the plasma deviates in different directions from the centre when approaching disruption on EAST and J-TEXT. In the previous research, the interpretable study of J-TEXT proves that the plasma usually tends to shift towards the low field side (LFS) when approaching disruption (even if it is salvageable). However, the plasma tends to shift towards the high-field side (HFS) approaching disruption on EAST. This might be related to the differences in plasma control systems between J-TEXT and EAST.

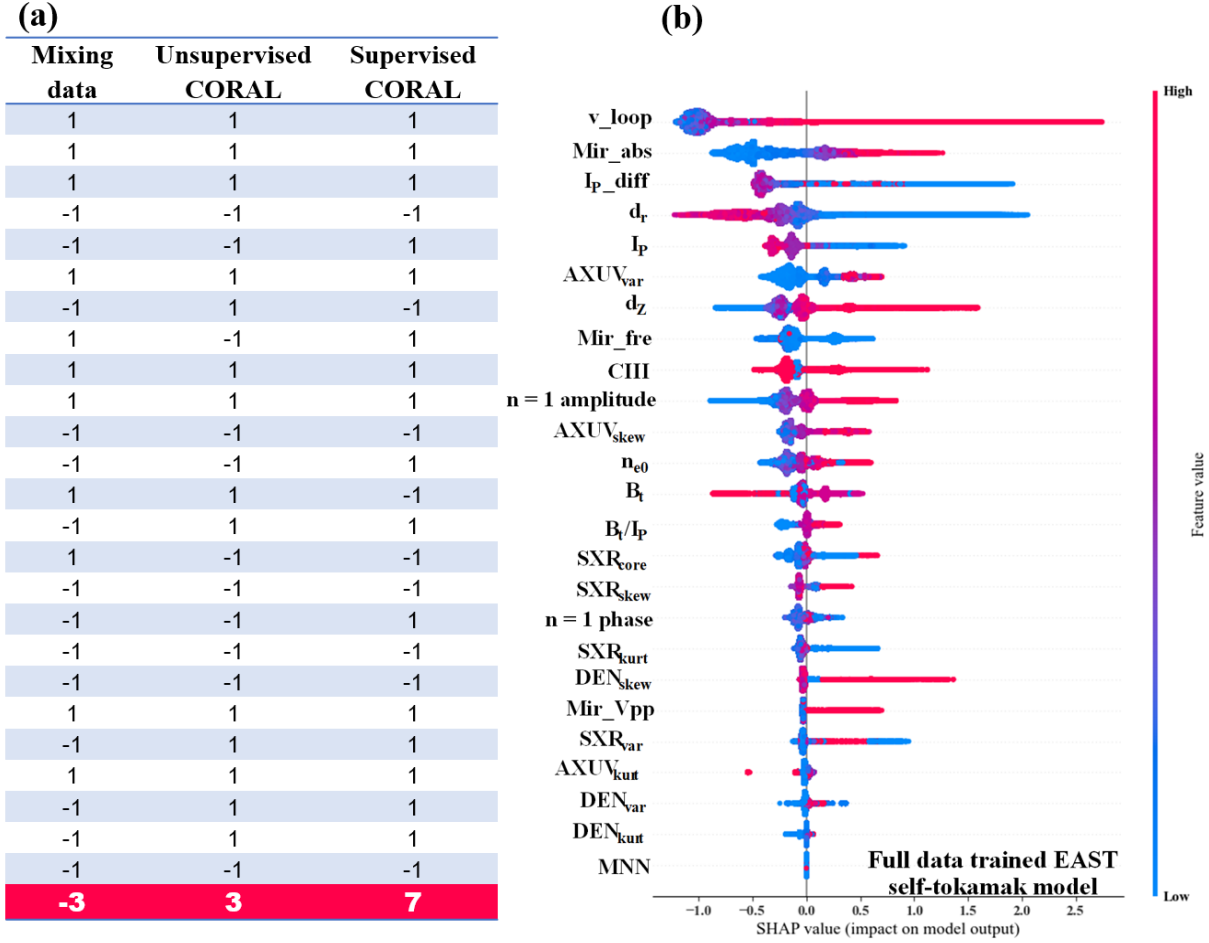

Figure 9 (a) The Scores of three cross-tokamak models by the similarity evaluation with the self-tokamak model. The last row of the table with the red background colour shows the total score. (b) The SHAP value of different features and their relations with feature value of the full data trained EAST self-tokamak model. The width of bar in the SHAP result represent the number of the samples. The larger the width of the bar, the larger the number of samples. The order of the features represents the contributions of features. The colormap represents the feature value of each feature, red means high and blue means low. A positive SHAP value represents a "disruptive" impact on the model, while a negative SHAP value represents a "non-disruptive" impact on the model.

## 6. Summary and future plan

This paper introduced a novel approach to predict disruption in a future tokamak only using a few discharges based on PGFE and CORAL. This approach is a light, interpretable and few data required cross-tokamak approach. It is the first attempt of applying domain adaptation in the task of disruption prediction.

Cross-tokamak disruption prediction based on PGFE and CORAL aligns a few data from the future tokamak (target domain) and large amount of data from existing tokamak (source domain) to train a machine learning model in the existing tokamak. We selected J-TEXT and EAST to simulate the existing and future tokamak, respectively. PGFE, originally designed as a feature extractor for J-TEXT, has now been successfully implemented on EAST. Moreover, it has achieved a high-performance EAST self-tokamak model (AUC = 0.936, TPR = 93.33%, FPR = 16.11%) using a large amount of data from EAST. This demonstrates that PGFE possesses the adaptability to be transferred to other tokamaks. PGFE can extract the less device-specific features, which established a solid foundation for cross-tokamak disruption prediction. However, difference in tokamaks might result in distinct decision boundaries on disruption. Therefore, CORAL as a domain adaptation algorithm is used to transfer the disruption prediction model from J-TEXT to EAST. In this paper, CORAL is improved into an algorithm that is more suitable for the disruption prediction task, call supervised CORAL (S-CORAL). The original CORAL, on the other hand, is referred to as unsupervised CORAL (U-CORAL) in this paper. With limited EAST data (100 non-disruptive discharges and 10 disruptive discharges), the commonly used mixing data method fails to achieve good performance (AUC = 0.764, TPR = 73.33%, FPR = 27.78%) only using PGFE to align features. Using U-CORAL can enhance the performance of disruption prediction on EAST with the TPR of 90.56%, FPR of 46.67% and AUC value of 0.797. Using S-CORAL further improves the disruption prediction performance on future tokamak with the TPR of 90%, FPR of 25.56% and AUC value of 0.89. The interpretability study shows the reason that why S-CORAL model could perform the best in the three cross-tokamak models in this paper. From the analysis of the data distribution, S-CORAL brings the transformation of the data distribution closer to EAST than to J-TEXT. Moreover, SHAP analysis was done on both the EAST self-tokamak model as well as all three cross-tokamak models. We propose an assessment method for evaluating whether a model has learned a trend of similar features using SHAP analysis. It is found that the S-CORAL model (scores 7) learned knowledge more similar to the EAST self-tokamak model than other two models (mixing data model scores -3 and unsupervised CORAL model scores 3). Based on the SHAP analysis, we hypothesize that differences in the control systems of different tokamaks may affect the transfer effects of the disruption prediction models.

Although this paper proposes a light, interpretable and few data required cross-tokamak approach, it still need to be improved. (1) Only the J-TEXT and EAST are used to test this cross-tokamak disruption prediction approach. Data from more tokamak would be beneficial for validation and improvement of this approach. (2) PGFE still could not extract generalized normalized features, although it has been successfully applied on EAST. Therefore, the improvement of PGFE applicable to most tokamaks requires continuous and in-depth research. (3) The performance of self-tokamak model still needs to be improved. The FPR of the model needs to be further reduced to ensure the economics of future tokamak operations. High FPR can cause a significant reduction in discharge efficiency. (4) PGFE is not only a feature alignment algorithm for decision tree, but also could be applied in deep learning. (5) The cross-tokamak disruption prediction models should require fewer data from the future tokamak (target domain), such as zero-shot test. Therefore, our team would like to first improve PGFE and try to applied on other tokamaks. We will also explore possible cross-tokamak disruption prediction approaches with fewer data from the future tokamak.

## Acknowledgement

This work was supported by National Key R&D Program of China under Grant (No. 2022YFE03040004 and No. 2019YFE03010004) and by National Natural Science Foundation of China (NSFC) under Project Numbers Grant (No. 12075096 and No. 51821005).

## Data availability

Raw data were generated at the J-TEXT and EAST facilities. Derived data are available from the corresponding author upon reasonable request.

## Code availability

The computer code that was used to generate figures and analyze the data is available from the corresponding author upon reasonable request.